\documentclass[twocolumn]{aastex631}

\usepackage{hyperref}
\usepackage{color}

\newcommand{\kms}{km s$^{-1}$}

\definecolor{notes}{HTML}{5C9384}

\newcommand{\oiiidbl}{[\textrm{O}\textsc{iii}]\ensuremath{\lambda\lambda4960,5008}}
\newcommand{\oiiiauroral}{[\textrm{O}\textsc{iii}]\ensuremath{\lambda4363}}

\newcommand{\oiibrightdbl}{[\textrm{O}\textsc{ii}]\ensuremath{\lambda\lambda3727,3730}}
\newcommand{\oiiauroraldbl}{[\textrm{O}\textsc{ii}]\ensuremath{\lambda\lambda7320,7330}}

\newcommand{\siidbl}{[\textrm{S}\textsc{ii}]\ensuremath{\lambda\lambda6718,6733}}
\newcommand{\siiiauroral}{[\textrm{S}\textsc{iii}]\ensuremath{\lambda6314}}
\newcommand{\siiauroraldbl}{[\textrm{S}\textsc{ii}]\ensuremath{\lambda\lambda4070,4076}}

\newcommand{\siiib}{[\textrm{S}\textsc{iii}]\ensuremath{\lambda9532}}

\newcommand{\niiauroral}{[\textrm{N}\textsc{ii}]\ensuremath{\lambda5756}}
\newcommand{\niidbl}{[\textrm{N}\textsc{ii}]\ensuremath{\lambda\lambda6550,6585}}

\newcommand{\feiib}{[\textrm{Fe}\textsc{ii}]\ensuremath{\lambda4288}}
\newcommand{\feii}{[\textrm{Fe}\textsc{ii}]\ensuremath{\lambda4360}}
\newcommand{\feiii}{[\textrm{Fe}\textsc{iii}]\ensuremath{\lambda4658}}

\newcommand{\ariii}{[\textrm{Ar}\textsc{iii}]\ensuremath{\lambda7138}}
\newcommand{\arivdbl}{[\textrm{Ar}\textsc{iv}]\ensuremath{\lambda\lambda4713,4741}}
\newcommand{\ariva}{[\textrm{Ar}\textsc{iv}]\ensuremath{\lambda4713}}

\newcommand{\neiiiauroral}{[\textrm{Ne}\textsc{iii}]\ensuremath{\lambda3343}}

\newcommand{\halpha}{\textrm{H}\ensuremath{\alpha}}
\newcommand{\hbeta}{\textrm{H}\ensuremath{\beta}}
\newcommand{\hgamma}{\textrm{H}\ensuremath{\gamma}}
\newcommand{\hdelta}{\textrm{H}\ensuremath{\delta}}

\newcommand{\heiiopt}{\textrm{He}\textsc{ii}\ensuremath{\lambda4686}}
\newcommand{\heia}{\textrm{He}\textsc{i}\ensuremath{\lambda4471}}

\newcommand{\siitemp}{[\textrm{S}\textsc{ii}]\ensuremath{\lambda4069/\lambda\lambda6718,6733}}
\newcommand{\siiitemp}{[\textrm{S}\textsc{iii}]\ensuremath{\lambda6314/\lambda9071}}

\newcommand{\niitemp}{[\textrm{N}\textsc{ii}]\ensuremath{\lambda5756/\lambda6585}}

\begin{document}

\title{The Sunburst Arc with JWST: III. An Abundance of Direct Chemical Abundances}

\correspondingauthor{Brian Welch}
\email{brian.d.welch@nasa.gov}

\author[0000-0003-1815-0114]{Brian Welch}
\affiliation{Department of Astronomy, University of Maryland, College Park, MD 20742, USA} 
\affiliation{Astrophysics Science Division, Code 660, NASA Goddard Space Flight Center, 8800 Greenbelt Rd., Greenbelt, MD 20771, USA}
\affiliation{Center for Research and Exploration in Space Science and Technology, NASA/GSFC, Greenbelt, MD 20771}

\author[0000-0002-9204-3256]{T. Emil Rivera-Thorsen}
\affiliation{The Oskar Klein Centre, Department of Astronomy, Stockholm University, AlbaNova 10691, Stockholm, Sweden}

\author[0000-0002-7627-6551]{Jane R. Rigby}
\affiliation{Astrophysics Science Division, Code 660, NASA Goddard Space Flight Center, 8800 Greenbelt Rd., Greenbelt, MD 20771, USA}

\author[0000-0001-6251-4988]{Taylor A. Hutchison}
\affiliation{Astrophysics Science Division, Code 660, NASA Goddard Space Flight Center, 8800 Greenbelt Rd., Greenbelt, MD 20771, USA}

\author[0000-0002-4606-4240]{Grace M. Olivier}
\affiliation{Department of Physics and Astronomy and George P. and Cynthia Woods Mitchell Institute for Fundamental Physics and Astronomy, Texas A\&M University, 4242 TAMU, College Station, TX 77843-4242, USA}

\author[0000-0002-4153-053X]{Danielle A. Berg}
\affiliation{Department of Astronomy, The University of Texas at Austin, 2515 Speedway, Stop C1400, Austin, TX 78712, USA}

\author[0000-0002-7559-0864]{Keren Sharon}
\affiliation{Department of Astronomy, University of Michigan, 1085 S. University Ave, Ann Arbor, MI 48109, USA}

\author[0000-0003-2200-5606]{H\r{a}kon Dahle} \affiliation{Institute of Theoretical Astrophysics, University of Oslo, P.O. Box 1029, Blindern, NO-0315 Oslo, Norway}

\author[0000-0002-2862-307X]{M. Riley Owens}
\affiliation{Department of Physics, University of Cincinnati, Cincinnati, OH 45221, USA}

\author[0000-0003-1074-4807]{Matthew B. Bayliss}
\affiliation{Department of Physics, University of Cincinnati, Cincinnati, OH 45221, USA}

\author[0000-0002-3475-7648]{Gourav Khullar}
\affiliation{Department of Physics and Astronomy and PITT PACC, University of Pittsburgh, Pittsburgh, PA 15260, USA}

\author[0000-0002-0302-2577]{John Chisholm}
\affiliation{Department of Astronomy, The University of Texas at Austin, 2515 Speedway, Stop C1400, Austin, TX 78712, USA}

\author[0000-0001-8587-218X]{Matthew Hayes}
\affiliation{The Oskar Klein Centre, Department of Astronomy, Stockholm University, AlbaNova 10691, Stockholm, Sweden}

\author[0000-0001-6505-0293]{Keunho J. Kim}
\affiliation{IPAC, California Institute of Technology, 1200 E. California Boulevard, Pasadena, CA 91125, USA}

\begin{abstract}

We measure the gas-phase abundances of the elements He, N, O, Ne, S, Ar, and Fe in an individual H\textsc{ii} region known to be leaking Lyman-continuum photons in the Sunburst Arc, a highly magnified galaxy at redshift $z=2.37$. 
We detect the temperature-sensitive auroral lines \siiauroraldbl, \oiiiauroral, \siiiauroral, \oiiauroraldbl, and \neiiiauroral ~ in a stacked spectrum of 5 multiple images of the Lyman-continuum emitter (LCE), from which we directly measure the electron temperature in the low, intermediate, and high ionization zones. 
We also detect the density-sensitive doublets of \oiibrightdbl, \siidbl, and \arivdbl, which constrain the density in both the low- and high-ionization gas.
With these temperature and density measurements, we measure gas-phase abundances with similar rigor as studies of local galaxies and H\textsc{ii} regions.
We measure a gas-phase metallicity for the LCE of $12+\log(\textrm{O}/\textrm{H}) = 7.97 \pm 0.05$, and find an enhanced nitrogen abundance $\log(\textrm{N}/\textrm{O}) = -0.65^{+0.16}_{-0.25}$.
This nitrogen abundance is consistent with enrichment from a population of Wolf-Rayet stars, additional signatures of which are reported in a companion paper. 
Abundances of sulfur, argon, neon, and iron are consistent with local low-metallicity H\textsc{ii} regions and low-redshift galaxies. 
This study represents the most complete chemical abundance analysis of an individual H\textsc{ii} region at Cosmic Noon to date, which enables direct comparisons between local H\textsc{ii} regions and those in the distant universe.

\end{abstract}


\section{Introduction}\label{sec:intro}

As stars shed mass in their late evolutionary stages and explode as supernovae, they recycle newly synthesized elements into the surrounding gas, enriching the nebula. 
Analysis of the chemical abundance patterns in nebulae thus gives insight into the evolution of stars and the star formation histories of galaxies. 

Recent measurements of high redshift ($z>8$) galaxies have found unusual abundance patterns, including anomalously high nitrogen abundances \citep[N/O, ][]{Cameron23,Senchyna23,Marquez-Chavez24,Castellano24} and carbon abundances \citep[C/O, ][]{DEugenio23}. 
Meanwhile, similar carbon abundance measurements at slightly lower redshifts ($z=4-6$) find much lower abundances \citep{Citro23,Jones23}.
The anomalous abundance patterns in the highest redshift galaxies may be indicative of chemical evolution via e.g. Wolf-Rayet (WR) stars or extremely metal poor stars, as the standard production pathway for nitrogen and carbon via intermediate-mass AGB stars would not have time to develop.  

One of the most robust methods to determine chemical abundance patterns in galaxies is the so-called ``direct" method, which directly measures electron temperature ($T_e$) and density from the strengths of auroral emission lines relative to bright collisionally excited lines.
These temperatures and densities are then used in conjunction with the strengths of the bright lines to calculate the abundance of each ion within the nebula \citep{Dinerstein1990}.

The auroral lines are intrinsically faint, making this measurement difficult.
Direct $T_e$ abundances have been well studied in nearby galaxies \citep[e.g.,][]{VanZee_Haynes2006,Berg15,Croxall15,Croxall16,Berg20,Rogers21,Rogers22}.
In these local galaxies, multiple auroral lines are detected, allowing direct measurements of the electron temperature in multiple ionization zones within the nebular gas. 
This eliminates the need for empirical temperature relations to infer the physical conditions of different zones within the gas, thereby making the final abundance determinations more accurate. 
Additionally, including multiple ionization zones minimizes the impact of empirically calibrated ionization corrections, which account for unobserved ionization states of a given element \citep[e.g., ][]{Izotov06,Amayo21}.
In the best case, all relevant ionization states of each element are observed, resulting in the most accurate determinations of elemental abundances.

Only a handful of auroral line detections were available for galaxies at higher redshifts ($z\sim 1-3$) prior to the launch of JWST \citep{Christensen12,James14,Bayliss14,Ly15,Kojima17,Sanders2020,Sanders23_keck,Citro23}.
The sensitivity and wavelength coverage of JWST \citep{Gardner23_jwst,Rigby23_jwstperf} have changed the landscape of direct $T_e$ chemical abundance measurements in distant galaxies. 
The auroral \oiiiauroral ~line was detected out to $z \sim 8$ for the first time in the first data release from the Early Release Observation \citep[ERO, ][]{Pontoppidan22_ERO} of the lensing cluster SMACS0723 \citep{ArellanoCordova22,Schaerer22,Taylor22,Brinchmann23,Curti23auroral,Katz23,Rhoads23,Trump23,Trussler23}.
This line has since been clearly detected out to $z\sim 8.7$ in a field galaxy \citep{Sanders23_ceers}, and at $z = 10.17$ in a gravitationally lensed galaxy \citep{Hsiao24}.

Each of the aforementioned studies have relied on detection of a single auroral line for their abundance measurements.
Several early JWST observing programs have detected multiple auroral lines in galaxies at Cosmic Noon, enabling more robust studies of abundance patterns in distant galaxies. 
The CECILIA program (PID 2593, PI Strom, Co-PI Rudie) targeted the \niiauroral, \siiiauroral, and \oiiauroraldbl ~lines in a sample of galaxies at $z\sim 2-3$, aiming to measure multiple ionization states and thus obtain more accurate abundances. This program has so far yielded an analysis of the stacked sample \citep{Strom23_Cecilia_lines}, and measurements of all three lines in a single galaxy at $z\sim3$ with a notably sub-solar S and Ar abundance \citep{Rogers24}. 
Meanwhile the TEMPLATES ERS program (PID 1355, PI Rigby, Co-PI Viera) has detected \oiiiauroral, \siiiauroral, and \oiiauroraldbl ~in a lensed galaxy at $z\sim1.3$, finding a slightly enhanced nitrogen abundance \citep{Welch24}.
These studies measuring multiple ionization states at $z\sim 1-3$ represent a positive step towards applying the same detailed methods used in local galaxies to high redshift. 
However none have managed to reach the complete standard set by studies of local galaxies.

Measurements of direct $T_e$ abundances in distant galaxies have thus far only been made using the integrated light of an entire galaxy. Studies of nearby galaxies have the advantage of resolving individual H\textsc{ii} regions within galaxies and measuring the abundances for these regions independently, finding that abundances tend to vary across galactocentric radius \citep[e.g.,][]{Zaritsky94,Moustakas2010,SanchezMenguiano18,Berg20}. Abundance measurements using galaxy-integrated spectra therefore do not provide a complete picture of the abundance structure within the galaxy.

Abundance patterns of sub-galactic objects can also shed light on their growth and development. For example, multiple stellar populations found in globular clusters show clear abundance trends between the first and second generations of stars \citep[e.g.,][]{BastianLardo2018}. The exact mechanism of enrichment for second generation stars is debated, with fast rotating massive stars, AGB stars, binary interactions, and supermassive stars (among others) proposed as the sources of enrichment (see e.g. \cite{Charbonnel16rev} for a review). Measurements of individual H\textsc{ii} regions and proto-globular cluster candidates in distant galaxies can help to shed additional light on the questions of chemical evolution at the sub-galactic scale and the growth of globular clusters.

This paper presents the analysis of direct $T_e$ abundances in the Sunburst Arc \citep{Dahle16}, a highly magnified gravitationally lensed galaxy at redshift $z=2.37$ \citep{Sharon22sunburst,Diego22,Pignataro21}. 
We focus on a single physical region of the galaxy that is leaking ionizing photons \citep{RivThor19}, which we refer to as the Lyman-continuum emitter (LCE). 
This region is a compact star cluster \citep{Vanzella22} hosting a young stellar population with a light-weighted age of $\sim 3.6$ Myr \citep{Chisholm19,RivThor24} and a steep UV slope $\beta \simeq -3$ \citep{Kim23}. Several authors have found evidence of very massive stars in the LCE \citep{Chisholm19,Mestric23,Pascale23}.
A companion paper presents evidence that the LCE hosts a population of Wolf-Rayet (WR) stars \citep{RivThor24}. 

In this paper, we report the detection of the \siiauroraldbl, \oiiiauroral, \siiiauroral, \oiiauroraldbl, and \neiiiauroral ~auroral lines, from which we measure $T_e$ directly in the low, intermediate, and high ionization zones of the nebula. 
We also measure the electron temperature in ionized hydrogen gas from the Balmer jump.
We measure abundances for seven elements (He, N, O, Ne, S, Ar, Fe).
For three of these (O, S, Ar) we measure multiple ionization states, which minimizes the impact of ionization corrections in our abundance determinations. 
With this set of diagnostics, for the first time we apply the same level of rigor used in local H\textsc{ii} regions to an object of similar physical scale at Cosmic Noon, bridging the gap between local abundance measurements and those at high redshift. 

This paper is organized as follows.
The data are presented in Section \ref{sec:data}. 
Our emission line measurements and reddening corrections are described in Section \ref{sec:methods}.
We describe our electron temperature, density, and chemical abundance measurements and results in Section \ref{sec:abunds}, and discuss the context of these results in Section \ref{sec:discussion}. 
We summarize our conclusions in Section \ref{sec:conclusion}.
We assume flat $\Lambda$CDM cosmology with 
$\Omega_{\Lambda} = 0.7$, $\Omega_{m}=0.3$, and $H_0 = 70$ \kms\ Mpc$^{-1}$.

\section{Data} \label{sec:data}

The Sunburst Arc was observed by JWST in Cycle~1 using NIRCam \citep{Rieke23_nircam} and the Integral Field Spectroscopy (IFS) mode of NIRSpec \citep{Boker23_nirspec}, taken in program GO-02555 (PI: Rivera-Thorsen).
NIRCam data were taken in six broadband filters (F115W, F150W, F200W, F277W, F356W, and F444W). 
NIRSpec data were taken with two grating settings (G140H/F100LP, G235H/F170LP).
The IFS data were taken in three pointings, shown in Figure 1 of \cite{RivThor24}. 
The observations and data reduction are described in greater detail in Rivera-Thorsen et al., in prep; we briefly summarize the relevant steps here. 
The NIRSpec IFS data were reduced following the methods described in \cite{Rigby23_overview}, using the TEMPLATES NIRSpec data reduction notebook \citep{templates_data_notebooks}. 
We used the JWST data reduction pipeline version 1.11.4 \citep{Bushouse23_pipeine1p11p4}, along with calibration reference files from \texttt{pmap\_1105}.
After running the main JWST pipeline, we use the \texttt{baryon-sweep}\footnote{\href{https://github.com/aibhleog/baryon-sweep}{https://github.com/aibhleog/baryon-sweep},\\ DOI: \href{10.5281/zenodo.8377531}{10.5281/zenodo.8377531}} code to remove any remaining outliers from the final data cubes \citep{Hutchison23_sigmaclip,baryonsweep}.

In this work, we analyze a spatially-integrated spectrum that consists of 5 multiple images of the LCE covered by the three IFU pointings. 
We extract spectra of the LCE images from $5\times5$ spaxel apertures, using a flux-weighted sum. 
The spectra are normalized by the median flux in the wavelength region overlapping between the two gratings. No magnification correction is applied, however all results presented herein rely exclusively on flux ratios which are unaffected by magnification or normalization. 
Normalized spectra from the individual apertures are summed to create the final stacked spectrum.
The full extraction method is documented in greater detail in \cite{RivThor24}.

\section{Methods}\label{sec:methods}

\subsection{Continuum Subtraction}

Prior to fitting emission lines, we first remove the continuum. 
We mask known emission lines and sigma-clip the remaining continuum to remove any artifacts that passed our previous artifact removal steps described in Section \ref{sec:data}. The remaining continuum is then smoothed with a running median to produce a smooth continuum model. The smooth model is subtracted from the original spectrum prior to fitting any emission lines. 
This method of continuum subtraction will not account for features such as Balmer absorption. However, we do not expect significant Balmer absorption in the young stellar population seen in the LCE \citep[$\lesssim 4$ Myr,][]{Chisholm19,RivThor24}, and visual inspection of the original spectrum shows no sign of Balmer absorption features. 

\subsection{Emission Line Measurements}

Emission lines are fit using Gaussian profiles, utilizing the publicly-available tools developed by the TEMPLATES team and available on GitHub.\footnote{\href{https://github.com/JWST-Templates}{github.com/JWST-Templates}}
We initially attempt to fit each line with a single Gaussian component, and iteratively add up to two additional components if visual inspection determines the fit to be poor. 
We find that the brightest lines (e.g. \oiiidbl, \halpha) require three components - one narrow (unresolved) component, one medium component with a full-width at half maximum (FWHM) of $\sim200-300$~km s$^{-1}$, and one broad component with a FWHM of $\sim 600-700$~km s$^{-1}$. 
The width of the medium component is consistent with measurements from \cite{Mainali22}, and all components are consistent with \cite{RivThor24}.
As line strengths decrease, the broadest component is the first to become lost in the noise. Thus for many lines, only the unresolved and the medium width components are used in fitting.
For the faintest lines (including all of the auroral lines), only a single unresolved component is sufficient to obtain a good fit. 
When fitting both single- and multi-component lines, we allow the centroids of each component to vary within $\pm 2$ \AA ~in the observed frame to allow for possible velocity offsets from our assumed systemic redshift \citep[$z=2.371062 \pm 6\times10^{-6}$][]{RivThor24}. 
For bright multi-component lines, the amplitude and width of the Gaussians are left unconstrained. 
For the faintest emission features, we choose to fix the line widths to the instrumental resolution reported in the JDox dispersion files\footnote{\href{https://jwst-docs.stsci.edu/jwst-near-infrared-spectrograph/nirspec-instrumentation/nirspec-dispersers-and-filters}{https://jwst-docs.stsci.edu/jwst-near-infrared-spectrograph/nirspec-instrumentation/nirspec-dispersers-and-filters}} at the observed wavelength, as is observed in the narrow components of the brighter emission lines. This prevents noise around the edge of the lines from artificially broadening the Gaussian fits. 
However we note that the NIRSpec IFU dispersions have yet to calibrated based on in-flight measurements, so it is possible that this method introduces a small systematic uncertainty.
Final line fluxes are obtained  by integrating the Gaussian fits. 
For line fits with two components, we sum the Gaussian integrals to obtain the total line flux. For three component fits, we choose to not include the broadest component in our total flux measurements, as the velocities ($\sim 600-700$~km s$^{-1}$) are such that this gas has likely traveled far from the main nebula. This only affects the \halpha ~and \oiiidbl ~lines.
Uncertainties in measured line fluxes are propagated through from fitting uncertainties on the Gaussian amplitude and width. 

\begin{figure}
    \centering
    \includegraphics[width=0.46\textwidth]{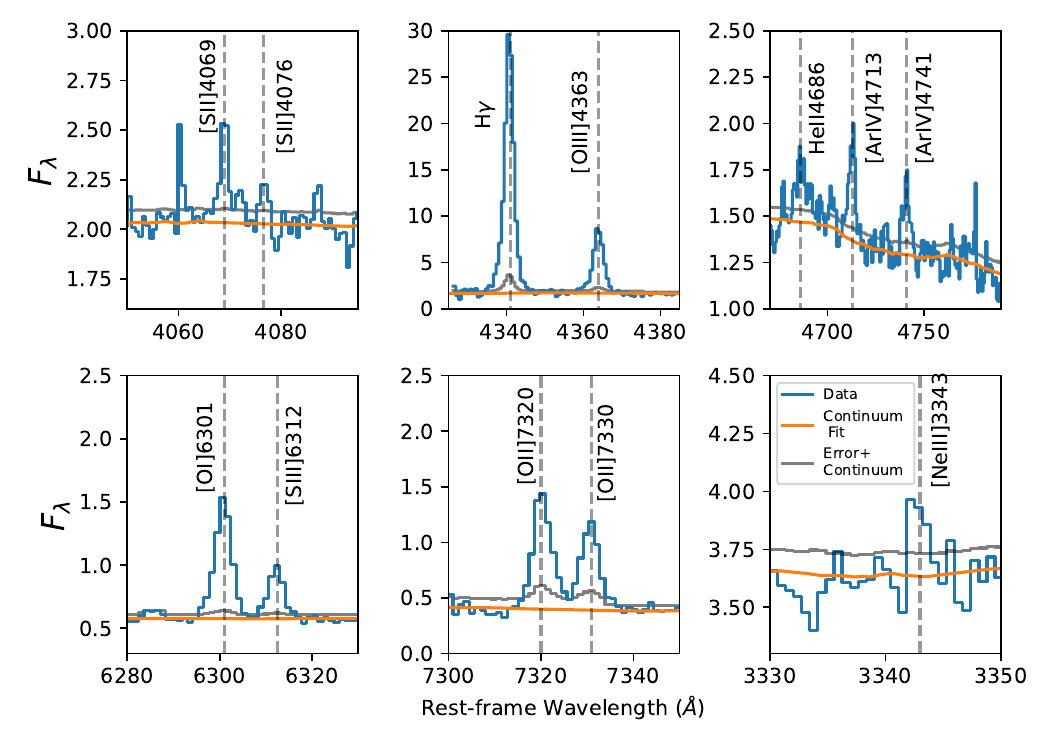}
    \caption{Stacked spectrum of 5 multiple images of the Lyman-continuum emitter in the Sunburst Arc, zoomed in to highlight detected auroral emission lines. The extracted spectrum is shown in blue, and the continuum fit is shown in orange. Flux errors are shown in grey, offset by the continuum fit to show the SNR of the continuum-subtracted lines. The \arivdbl~lines abut the blue Wolf-Rayet bump \citep{RivThor24}. Our continuum fit treats this broad feature similar to the continuum, so it is removed prior to fitting the argon emission lines.}
    \label{fig:spec}
\end{figure}

Blended emission lines (e.g., \halpha\ and \niidbl) are fit simultaneously, using the same priors and uncertainty calculations described above. 
One notable exception in our fitting process is the \oiibrightdbl~doublet. 
These lines are blended together, and because of the high signal-to-noise ratio (SNR) of this spectrum, the H13 and H14 lines of hydrogen are also blended with the \oiibrightdbl~doublet. 
We find that the \oiibrightdbl ~lines require two components each to fit properly. 
The combination of six blended Gaussian profiles leads to a large degeneracy in model parameters, particularly within the \oiibrightdbl ~doublet. 
We choose to fix the parameters of the H13 and H14 lines to improve the accuracy of the \oiibrightdbl~fit. The fluxes of H13 and H14 are calculated from our measured fluxes of the H11 and H12 lines, which are well separated from other nearby lines. We use a simple mean of the line fluxes predicted from H11 and H12 as our estimates for the H13 and H14 line flux. We fix the line widths of the H13 and H14 lines to be unresolved, and fix the observed wavelengths using the systemic redshift of the system. 
This process leads to a reliable \oiibrightdbl~fit, however the degeneracy between parameters in these blended lines still leads to unreasonably large flux uncertainties. 
To estimate the true uncertainty on the fluxes of these two lines, we run a MCMC ensemble using the python package \texttt{emcee} \citep{ForemanMackey13_emcee}. We calculate the total flux of each line from each step of the MCMC, and use the standard deviation of the resulting distribution as our estimate of the flux uncertainty.

The \ariva ~line is blended with a neighboring He\textsc{i} emission line at $4714$~\AA. To correct for the contribution from He\textsc{i}, we calculate the expected flux of this line based on the nearby \heia ~line, assuming a temperature of 15000 K and a density of 1200 cm$^{-2}$. The calculated strength of this He\textsc{i}$\lambda4714$ line is reported in Table \ref{tab:fluxes}, with uncertainty propagated from the \heia ~flux measurement. The calculated He\textsc{i}$\lambda4714$ is subtracted from the original measured \ariva ~line flux, and the uncertainties are added in quadrature to produce the final \ariva ~ flux reported in Table \ref{tab:fluxes}. 

The \niiauroral ~ line is not detected in our spectrum. We set a $3\sigma$ upper limit on this line flux using the summed flux in a neighboring region with no emission or absorption lines present. We randomly perturb the flux density in each wavelength step in this region by sampling a Gaussian distribution centered on the measured flux density, with the standard deviation set by the flux density uncertainty. We then sum the randomly perturbed flux in the region. We repeat this process 1000 times, then use the 99.85th percentile of the resulting summed flux distribution as our $3\sigma$ upper limit. The resulting upper limit is reported in Table \ref{tab:fluxes}.

We test our measured upper limit in two ways. First, we inject mock emission lines into the blank region used for the \niiauroral ~upper limit calculation, then attempt to fit these injected lines with a Gaussian profile. We find that a 3$\sigma$ measurement from the mock line matches our upper limit calculation. As a second consistency check, we calculate the upper limit using blank spectral regions near our other two faintest lines, \neiiiauroral ~ and \siiauroraldbl. We find that the upper limits for these lines are consistent with the measured $\sim 3\sigma$ fluxes for these lines. From these tests, we conclude that our upper limit is reliable.

\begin{table}
\centering
\caption{Emission Line Fluxes}
\label{tab:fluxes}
\begin{tabular}{c c c c}
\hline \hline
Line Name & Wavelength & $F(\lambda)/F(\textrm{H}\beta)$ & $I(\lambda)/I(\textrm{H}\beta)$ \\
\hline
[Ne~III]~3343 & $3343.50$ & $0.003 \pm 0.001$ & $0.003 \pm 0.001$  \\

[O~II]~3727 & $3727.09$ &$0.28 \pm 0.03$ & $0.29 \pm 0.03$  \\

[O~II]~3730 & $3729.88$ &$0.22 \pm 0.03$ & $0.22 \pm 0.03$  \\

[Ne~III]~3870 & $3869.86$ &$0.56 \pm 0.03$ & $0.58 \pm 0.03$  \\

[Ne~III]~3969 & $3968.59$ &$0.17 \pm 0.02$ & $0.18 \pm 0.03$  \\

H$\epsilon$ & $3971.20$ &$0.14 \pm 0.03$ & $0.15 \pm 0.03$  \\

[S~II]~4070 & $4069.75$ &$0.006 \pm 0.001$ & $0.006 \pm 0.001$  \\

S~II]~4076 & $4075.79$ &$0.002 \pm 0.001$ & $0.003 \pm 0.001$  \\

H$\delta$ & $4102.89$ &$0.24 \pm 0.02$ & $0.25 \pm 0.02$  \\

H$\gamma$ & $4341.68$ &$0.52 \pm 0.07$ & $0.53 \pm 0.07$  \\

[Fe~II]~4360 & $4359.59$ &$0.010 \pm 0.002$ & $0.010 \pm 0.002$  \\

[O~III]~4363 & $4364.44$ &$0.135 \pm 0.006$ & $0.138 \pm 0.006$  \\

He~I~4473 & $4472.70$ &$0.04 \pm 0.01$ & $0.04 \pm 0.01$  \\

[Fe~III]~4658 & $4659.35$ &$0.008 \pm 0.002$ & $0.008 \pm 0.002$  \\

[Ar~IV]~4713 & $4712.69$ &$0.008 \pm 0.003$ & $0.009 \pm 0.003$  \\

He~I~4714\tablenotemark{*} & $4714.47$ & $0.006 \pm 0.001$ & $0.007 \pm 0.001$ \\

[Ar~IV]~4741 & $4741.45$ &$0.010 \pm 0.002$ & $0.010 \pm 0.002$  \\

H$\beta$ & $4862.68$ &$1.0 \pm 0.1$ & $1.0 \pm 0.1$  \\

[O~III]~4960 & $4960.30$ &$2.2 \pm 0.3$ & $2.2 \pm 0.3$  \\

[O~III]~5008 & $5008.24$ &$6.8 \pm 0.4$ & $6.8 \pm 0.4$  \\

[N~II]~5756 & $5756.24$ & $<0.001$ & $<0.001$  \\

He~I~5878 & $5877.59$ &$0.142 \pm 0.007$ & $0.139 \pm 0.007$  \\

[O~I]~6302 & $6302.05$ &$0.026 \pm 0.004$ & $0.025 \pm 0.004$  \\

[S~III]~6314 & $6313.80$ &$0.010 \pm 0.001$ & $0.010 \pm 0.001$  \\

[N~II]~6550 & $6549.85$ &$0.040 \pm 0.004$ & $0.039 \pm 0.004$  \\

H$\alpha$ & $6564.61$ &$3.0 \pm 0.3$ & $2.9 \pm 0.3$  \\

[N~II]~6585 & $6585.28$ &$0.113 \pm 0.005$ & $0.109 \pm 0.005$  \\

He~I~6679 & $6680.00$ &$0.04 \pm 0.02$ & $0.04 \pm 0.02$  \\

[S~II]~6718 & $6718.29$ &$0.040 \pm 0.003$ & $0.039 \pm 0.003$  \\

[S~II]~6733 & $6732.67$ &$0.044 \pm 0.003$ & $0.043 \pm 0.003$  \\

He~I~7064 & $7064.21$ &$0.08 \pm 0.01$ & $0.08 \pm 0.01$  \\

[Ar~III]~7138 & $7137.80$ &$0.062 \pm 0.007$ & $0.059 \pm 0.006$  \\

[O~II]~7322 & $7322.01$ &$0.031 \pm 0.002$ & $0.030 \pm 0.002$  \\

[O~II]~7332 & $7331.68$ &$0.023 \pm 0.001$ & $0.022 \pm 0.001$  \\

[Ar~III]~7753 & $7753.20$ &$0.015 \pm 0.001$ & $0.014 \pm 0.001$  \\

[S~III]~9071 & $9071.10$ &$0.151 \pm 0.007$ & $0.139 \pm 0.007$  \\

\hline

\end{tabular}
\tablecomments{Emission line fluxes and dereddened intensities relative to \hbeta. Columns are: (1) Line identification, (2) Vacuum wavelength, in \AA, (3) Measured flux relative to \hbeta, (4) Dereddened intensity relative to \hbeta. 
Line fluxes were measured from the stacked spectrum of multiple images of the LCE, which has arbitrary units; to enable conversion to physical units, we report the measured and the demagnified flux of \hbeta ~for the image of the LCE that appears in image 4 of the lensed galaxy: the measured flux is $1.01 \pm 0.03 \times 10^{-17} \textrm{erg s}^{-1}\textrm{cm}^{-1}$, and the demagnified flux is $6.6 \pm 0.2 \times 10^{-19} \textrm{erg s}^{-1}\textrm{cm}^{-1}$, assuming the calculated magnification of $\mu = 15.3$ from \citet{Sharon22sunburst}.
Note this demagnified flux does not include magnification uncertainties.
}
\tablenotetext{*}{Calculated from He I$\lambda$4473 line flux}
\end{table} 

\subsection{Extinction Correction}

Emission line fluxes are first corrected for Milky Way dust reddening using the dust law of \cite{Cardelli1989}. 
We query the Milky Way dust map of \cite{SchlaflyFinkbeiner2011} via the NASA/IPAC Infrared Science Archive\footnote{\href{https://irsa.ipac.caltech.edu/applications/DUST/}{https://irsa.ipac.caltech.edu/applications/DUST/}} to obtain the ISM reddening correction. We find the Galactic $E(B-V) = 0.0812 \pm 0.003$.

We next correct for dust within the LCE itself. 
This correction is done using an iterative process using the \halpha, \hbeta, \hgamma, and \hdelta ~line strengths. 
We start by calculating the expected Balmer line ratios, assuming an initial electron temperature of $10^4$ K, and an initial electron density of $10^3$ cm$^{-3}$.
The measured line ratios are then compared to the predicted values to calculate $E(B-V)$, assuming the dust law of \cite{Cardelli1989}. 
We then apply the error-weighted average dust correction, weighted by the uncertainty on the Balmer line ratios, to relevant emission lines, and calculate the electron temperatures using the \oiiiauroral, \oiiauroraldbl, and \siiiauroral ~lines. 
The electron density is calculated using the \siidbl ~doublet ratio, as that is the best-measured of the density-sensitive doublets used in this study.
We update the electron temperature using an error-weighted average of the three calculated temperatures and repeat the dust correction measurement. 
This process is continued iteratively until the change in error-weighted electron temperature is less than 15 K.  The reddening that we calculated from this iterative process is $E(B-V) = 0.04 \pm 0.1$. The reddening-corrected line intensities are reported in  \autoref{tab:fluxes}.
Our measured reddening is consistent with that reported in \cite{RivThor24} using the attenuation law of \cite{Calzetti2000}, and with that reported from the modeling of \cite{Pascale23}. The reddening reported in \cite{Mainali22} is somewhat higher, however skylines impacting their \hbeta ~measurement could be the cause of the discrepancy.

As a consistency check, we also calculated the reddening correction assuming the dust law of \cite{Gordon03}, calculated for the Large Magellanic Cloud, to see if low metallicity alters the reddening result. We find  $E(B-V) = 0.06\pm 0.1$ in this case, consistent with the value calculated using \cite{Cardelli1989}.

\section{Direct Electron Temperatures and Chemical Abundances} \label{sec:abunds}

\subsection{Electron Temperature and Density}

The broad wavelength coverage (rest-frame $\sim 2900-9400$\AA) and high SNR of the stacked LCE spectrum enables calculations of direct electron temperatures and densities from multiple ions. 
We detect the auroral emission lines of \oiiiauroral, \oiiauroraldbl, \siiiauroral, \siiauroraldbl, and \neiiiauroral which are shown in Figure \ref{fig:spec}.
These let us use the temperature-sensitive auroral-to-nebular line ratios to calculate the electron temperatures for O$^{++}$, O$^+$, S$^{++}$, S$^+$, and Ne$^{++}$.
Our spectra do not go red enough to cover the \siiib ~line, so the S$^{++}$~temperature is measured from \siiitemp. 
While we detect both \siiauroraldbl~ lines, the redder line is barely detected ($< 3\sigma$). We therefore choose to use the ratio \siitemp~to calculate the temperature for this ion.
The \neiiiauroral ~line is weakly detected ($\sim 3\sigma$ significance), leading to a larger calculated uncertainty on the Ne$^{++}$ temperature.
We do not detect the \niiauroral ~line, and instead set a $3\sigma$ upper limit on its strength. This sets a corresponding $3\sigma$ upper limit on the ratio \niitemp, and thus an upper limit on the N$^+$ temperature. We also detect the density-sensitive lines \siidbl, \oiibrightdbl, and \arivdbl.
We calculate the electron temperatures and densities from these lines using the \texttt{getTemDen} function from the python package \texttt{PYNEB} \citep{pyneb_luridiana15}. 
The resulting measurements are presented in \autoref{tab:results}.

Uncertainties on temperatures and densities are calculated following the same method as \cite{Welch24}.
We draw 300 points from a Gaussian distribution centered on the measured line ratio, with standard deviation equal to the uncertainty on the line ratio. 
We calculate the temperature or density for each point, then take the standard deviation of the resulting temperature/density distribution as our uncertainty. 
In cases where the final distribution is highly asymmetric, we use the 16th and 84th percentiles as our $1\sigma$-low and $1\sigma$-high uncertainties. 

We calculate the electron temperatures assuming a fixed density. 
We use the best fit value from the \siidbl ~ratio ($n_e=1200$ cm$^{-3}$) as our fiducial density for temperature calculations. 
Similarly, we use a fixed temperature for our density calculations. 
We adopt the O$^{++}$ temperature ($T_e=15100$ K) as our fiducial value for density calculations.
The temperatures and densities are not strongly dependent on each other. 
We find that any reasonable change in density (within $\sim 1\sigma$ of the S$^+$ density) does not change our temperature estimates beyond the quoted uncertainty range, and any reasonable change in temperature (within $\sim 1\sigma$ of the O$^{++}$ temperature) does not change the density estimate beyond the quoted uncertainty range.

\subsubsection{Balmer Jump Temperature} \label{subsec:balmer}

\begin{figure}
    \centering
    \includegraphics[width=0.45\textwidth]{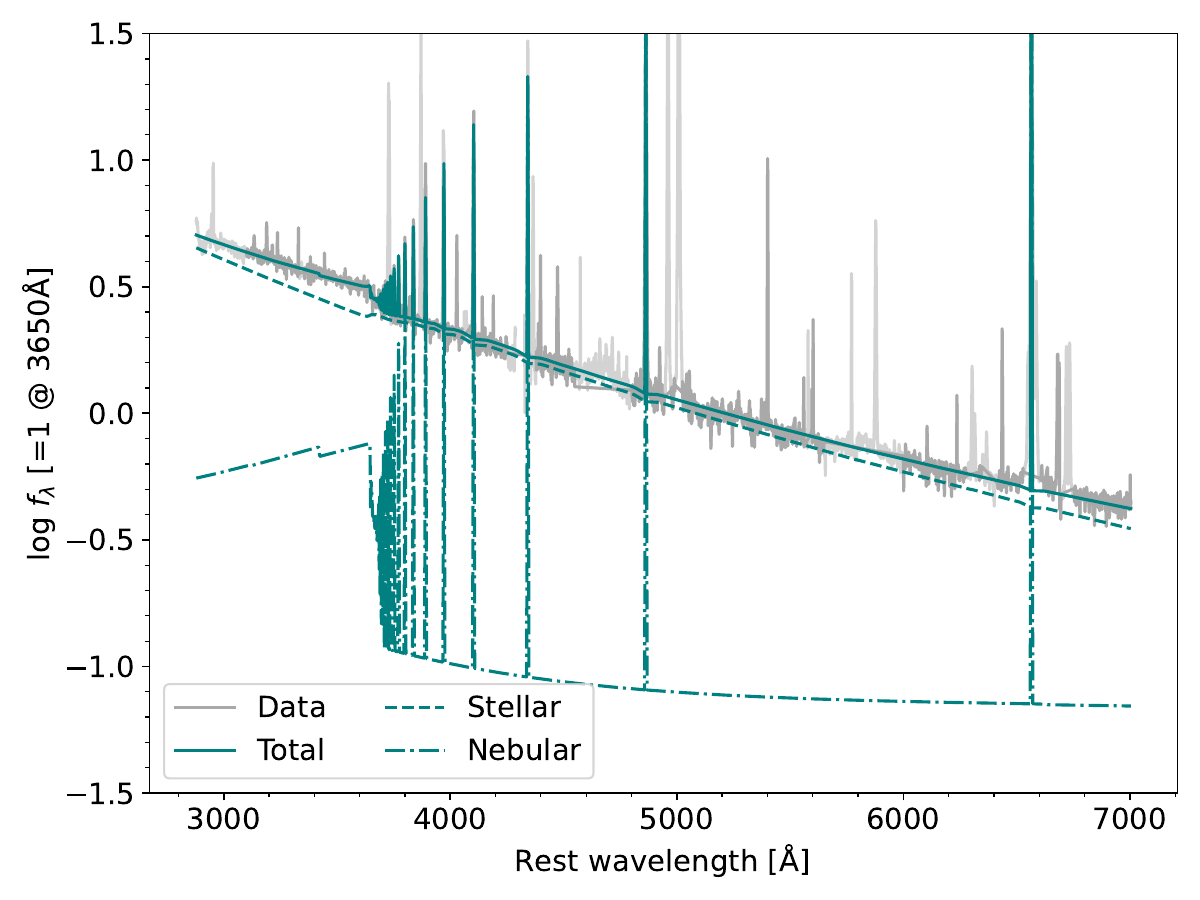}
    \caption{The spectrum of the Sunburst LCE shows a clear Balmer jump at 3646 \AA. The strength of this jump is sensitive to the electron temperature in the ionized hydrogen gas. We measure the strength of the Balmer jump by simultaneously fitting stellar and nebular models to our observed spectrum, masking our metal emission lines and WR features (see Section \ref{subsec:balmer} for details). The resulting best-fit model is shown in cyan, with the stellar and nebular components shown individually as dotted and dot-dashed lines respectively. The data are shown in grey. Our model successfully reproduces the observed strength of the Balmer jump. }
    \label{fig:balmerjump}
\end{figure}

Our spectrum covers the Balmer jump at $\lambda 3646$\AA, and the high SNR of the stacked spectrum allow us to detect this continuum feature.  
The strength of the jump is sensitive to the electron temperature of the ionized hydrogen in the nebula. 
The continuum around the Balmer jump feature is a combination of stellar and nebular emission, so we model these two components simultaneously following the procedure of \cite{Hayes24balmercalc}. The stellar component is based on Starburst99 template spectra \citep{Leitherer99,VasquezLeitherer05}, assuming a \cite{SalpeterIMF} IMF and a constant star formation history. The nebular spectra are created using the \cite{Schirmer16} code, and include contributions from recombination lines, free-bound, free-free, and 2-photon continuua. Both the stellar and nebular light are attenuated using the dust law of \cite{Calzetti2000}, though we find our results do not change significantly when using alternate dust attenuation prescriptions.

When fitting, we mask out metal emission lines, as these are not included in the model. We also mask the WR bumps reported in \cite{RivThor24}, as the Starburst99 stellar templates do not include WR emission. The region near the NIRSpec detector gap is also masked. Finally, the spectral templates used only extend to 7000\AA ~rest-frame, so we do not include the longer wavelengths in our fit. The data used, along with the best-fit stellar and nebular models, are shown in Figure \ref{fig:balmerjump}. 

This method yields a Balmer jump temperature of $8200 \pm 200$ K, consistent with our other measured low-ionization temperature ($T_e(\textrm{S}^+) = 10000 \pm 1000$ K) within $\sim 1.5\sigma$. As a consistency check for our model, we compare the age of the stellar population in our fit to the age measured in \cite{Chisholm19}. Our model finds a best-fit age of $2.9 \pm 0.6$ Myr, consistent with the $2.92 \pm 0.08$ Myr measured using a similar Starburst99 model in \cite{Chisholm19}.

We previously measured the Balmer jump following the procedure of \cite{Liu2001}, which provides a simple relation between Balmer jump strength and temperature assuming all the continuum is nebular in origin. 
With this assumption, we found a Balmer jump temperature of $T_e(\textrm{Bal}) = 12000 \pm 3000$ K. 
This temperature is higher than our model-based estimate, however the large uncertainty means that the two measurements are consistent within $\sim 1.3\sigma$.

\subsubsection{Comparison of Electron Temperatures \& Densities} \label{sec:tempdiscuss}

\begin{figure}
    \centering
    \includegraphics[width=0.45\textwidth]{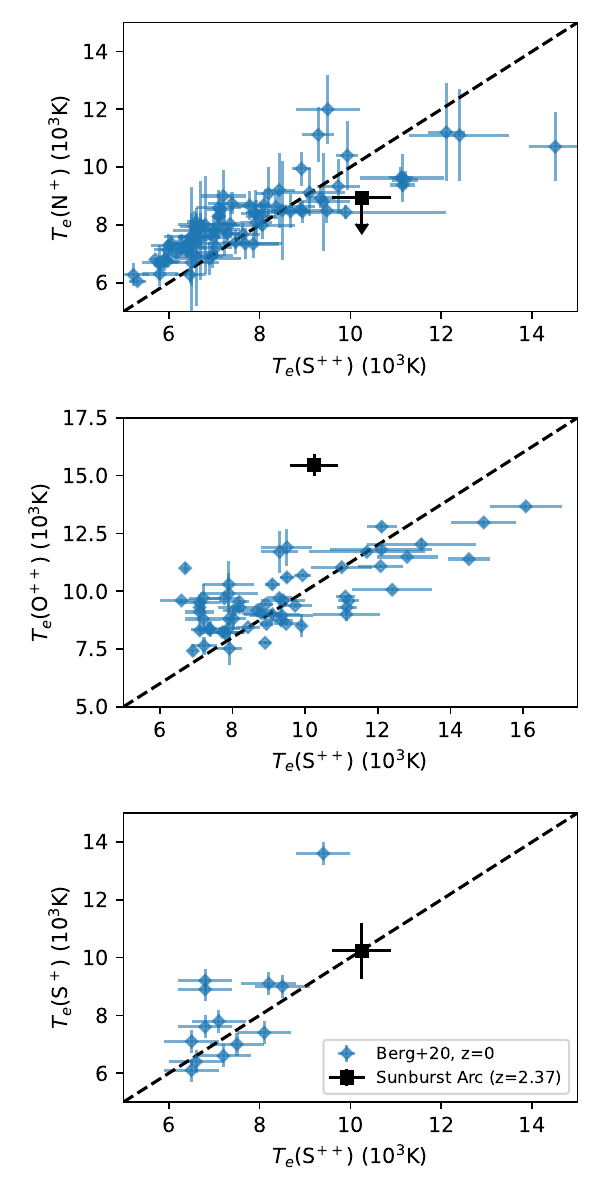}
    \caption{Comparison of  electron temperatures measured from auroral lines of different ions in the Sunburst Arc LCE (black point), plotted alongside local H\textsc{ii} regions from CHAOS \citep[blue points, ][]{Berg20}. The black dashed line represents equal temperatures. Our measured upper limit on the N$^+$ temperature is near the lower range of the local N$^+$ -- S$^+$ correlation (top panel), however it remains consistent with the local trend. The high-ionization O$^{++}$ temperature is significantly higher than the intermediate ionization S$^{++}$ temperature (middle panel), unlike what is seen in nearby H\textsc{ii} regions. Meanwhile the low-ionization and intermediate-ionization states of S are consistent with each other (lower panel). 
    }
    \label{fig:temps}
\end{figure}

We have measured electron temperatures and densities from multiple ionization zones within the Sunburst LCE nebula. 
The densities we measure are generally consistent within $1\sigma$ uncertainties, however there is some evidence that the high-ionization gas density is higher than the lower ionization states. 
While the low ionization estimates from S$^+$ ($n_e = 1200^{+500}_{-300} \textrm{cm}^{-3}$) and O$^+$ ($n_e = 1600^{+1500}_{-200}\textrm{cm}^{-3}$) are very consistent, the density measured from \arivdbl ~($n_e = 8000^{+10000}_{-5000}\textrm{cm}^{-3}$) is somewhat higher. 
However, the \arivdbl ~lines are much fainter than either \siidbl ~or \oiibrightdbl, leading to a much larger uncertainty on the resulting density estimate.
The high density reported from \arivdbl~ could thus be evidence that the high ionization gas is denser than the surrounding low-ionization gas, however the large uncertainty means that we cannot rule out these two ionization states having equal density. 

Electron temperature measurements show greater variation across ionization zones and across diagnostics. 
The lower-ionization gas probed by $T_e(\textrm{S}^+)$ (10.36--23.34 eV) and $T_e(\textrm{S}^{++})$ (23.34--34.86 eV) favors slightly lower electron temperatures ($10000\pm1000$ and $10300\pm600$ K, respectively). 
Meanwhile, the higher ionization states observed with $T_e(\textrm{O}^{++})$ (35.12--54.94 eV) and $T_e(\textrm{Ne}^{++})$ (40.96--63.42 eV) show slightly higher temperatures ($15100\pm500$ and $15400\pm1500$ K, respectively). 
These differences between ionization zones highlight the value of obtaining $T_e$ measurements from ions at different ionization energies, as empirical temperature relations do not fully capture such variations.

The differences in temperature between high-ionization and low- and intermediate-ionization gas are shown in Figure \ref{fig:temps}. The low and intermediate ionization states of sulfur have consistent temperatures, generally matching the trend seen in nearby H\textsc{ii} regions from the CHAOS sample \citep{Berg20}. The upper limit placed on the N$+$ temperature implies that the Sunburst LCE does not fit neatly with the local N$^+$ vs S$^{++}$ correlation, as the actual N$^+$ temperature would be low relative to $T_e(\textrm{S}^{++})$. However, \cite{Berg20} find increased scatter in this relation in high-ionization nebulae (determined by increased $F_{\lambda5008}/F_{\lambda3727,9}$), which is consistent with our finding for the highly ionized Sunburst LCE. Finally, we see that the high-ionization temperature of O$^{++}$ is significantly elevated relative to the local O$^{++}$ vs S$^{++}$ relation (middle panel of Figure \ref{fig:temps}). The local H\textsc{ii} regions from \cite{Berg20} do not include regions as highly ionized as the Sunburst LCE (their reported $F_{\lambda5008}/F_{\lambda3727,9}$ flux ratios only reach $\sim 1.6$, while the Sunburst LCE ratio is nearly an order of magnitude higher). In higher ionization galaxies, \cite{Berg21} find larger temperature gradients between the higher and lower ionization zones, similar to what we observe. Additionally, we find that the Ne$^{++}$ temperature is consistent with the O$^{++}$ temperature, corroborating that the high ionization gas is truly hotter than the lower ionization zones. We therefore conclude that the strong radiation sources within the stellar cluster powering the Sunburst LCE are driving the high ionization gas near the center of the nebula to higher temperatures than the lower-ionization gas towards the edge of the region.

One notable outlier in our measurements is $T_e(\textrm{O}^{+}) = 28000 \pm 3000$ K. 
At high densities ($n_e > 10^3\textrm{cm}^{-3}$), dielectronic recombination contributes substantially to the [OII] emission \citep{Liu2001}; this artificially increases the inferred O$^+$ electron temperature.
All of our densities are above this threshold, thus we conclude that the O$^+$ temperature is not reliable for this object. 
However, we note that the recombination contribution would have to be substantial to account for the full offset between our measured $T_e(\textrm{O}^+$) and the other low- and intermediate-ionization zone temperatures from S. It is therefore likely that other effects are contributing the this offset. One potential contributor is the nebular dust attenuation, since the wavelength gap between \oiibrightdbl ~ and \oiiauroraldbl ~ is large. Our measurement of dust attenuation assumes a \cite{Cardelli1989} dust law, however recently \cite{sanders24dust} found that the nebular attenuation curve for a $z\sim 4$ galaxy differed substantially from locally calibrated models. An altered attenuation curve would potentially bias the O$^+$ temperature that we measure. However such a dust attenuation curve measurement is beyond the scope of the current work.

The hydrogen gas temperature measured from the Balmer jump is lower than our other electron temperature measurements. The Balmer temperature is nearest to the other well-measured low-ionization temperature from S$^+$ ($T_e(\textrm{S}^+) = 10000 \pm 1000$ K); these two temperatures are consistent within $\sim 1.5\sigma$. The upper limit on $T_e(\textrm{N}^+) < 9000$ K is also similar to the Balmer temperature, though without a measured \niiauroral ~ line the N$^+$ temperature remains uncertain. The Balmer temperature is significantly lower than the measured high-ionization temperature from O$^{++}$. Interestingly, \cite{Hayes24balmercalc} find a similar discrepancy between the high-ionization temperature and the Bamer jump temperature. However, their stacked galaxy spectra show an anticorrelation between stellar population age and Balmer jump temperature, with which our measurement would appear inconsistent. Lastly, we note that \cite{Guseva06} find no significant discrepancies between temperatures measured from the Balmer jump and those from O$^{++}$. It is possible that the 5$\times$5 spaxel apertures used to create the spectrum analyzed here is large enough that the gas in the outskirts of the nebula biases the measured temperature low. Varying the size of the extraction aperture could test this hypothesis, however we leave this test for future work.

\subsection{Abundances and Ionization Corrections}

Ionic abundances relative to hydrogen are calculated using 
\begin{equation}
    \frac{N(X^i)}{N(H^+)} = \frac{I_{\lambda(i)}}{I_{\hbeta}} \frac{j_{\hbeta}}{j_{\lambda(i)}} .
\end{equation}
The emissivity coeffecients $j_{\lambda(i)}$ are temperature dependent. We calculate these coeffecients using the \texttt{getIonAbundance} function from \texttt{PYNEB} \citep{pyneb_luridiana15}. 
We use the temperature from the nearest ionization zones for each ionic abundance calculation, as described below. 

Atomic data used in our calculations are tabulated in Table \ref{tab:atomicdata}. 

\begin{table}[]
    \centering
    \scriptsize
    \caption{Atomic Data Sources}
    \label{tab:atomicdata}
    \begin{tabular*}{0.49\textwidth}{l l l}
    \hline \hline
        Ion & Transition Probabilities & Collision Strengths \\
    \hline 
        O$^+$ & \cite{FROESEFISCHER20041} & \cite{Kisielius09} \\
        O$^{++}$ & \cite{FROESEFISCHER20041} & \cite{Aggarwal99} \\
        N$^+$ & \cite{FROESEFISCHER20041} & \cite{Tayal11} \\
        S$^+$ & \cite{Rynkun19} & \cite{Tayal10} \\
        S$^{++}$ & \cite{Tayal19} & \cite{Grieve14} \\
        Ar$^{++}$ & \cite{Mendoza83} & \cite{MunozBurgos09} \\
        Ar$^{+3}$ & \cite{Rynkun19} & \cite{Ramsbottom97} \\
        Ne$^{++}$ & \cite{FROESEFISCHER20041} & \cite{McLaughlin2000} \\
    \hline
    \end{tabular*}
\end{table}

\subsubsection{Oxygen Abundance}

We calculate the total oxygen abundance $\textrm{O}/\textrm{H} = \textrm{O}^+/\textrm{H}^+ + \textrm{O}^{++}/\textrm{H}^+$. 
We use the electron temperature $T_e(\textrm{O}^{++})$ determined from \oiiiauroral ~ to calculate the $\textrm{O}^{++}/\textrm{H}^+$ ionic abundance.

For O$^+$, the electron temperature measured from the \oiiauroraldbl~lines is not reliable at the densities measured, as discussed in the Section \ref{sec:tempdiscuss}. 
We therefore use the S$^+$ temperature as our low-density electron temperature, as the \siiauroraldbl ~lines are not similarly inflated.
Calculating the $\textrm{O}^{+}/\textrm{H}^+$ abundance using the high O$^+$ temperature leads to a $\sim 10\times$ reduction in the inferred ionic abundance. 

The final oxygen ion abundances are reported in \autoref{tab:results}. We find a total metallicity for this object of $12+\log(\textrm{O}/\textrm{H}) = 7.97 \pm 0.05$.

\subsubsection{Nitrogen Abundance}

We make the assumption that N/O $\simeq \textrm{N}^+/\textrm{O}^+$, based on the similar ionization zones of the N$^+$ and O$^+$ ions \citep{Peimbert67}. 
This assumption has been found to be valid within $\sim 10\%$ \citep{Nava06,Amayo21}.

Because we do not detect the \niiauroral ~line, we use the temperature from the \siitemp ~ratio as our low-ionization temperature when calculating the nitrogen abundance. 
One potential caveat when using the S$^+$ temperature is that the lower bound of the S$^+$ ionization energy (10.36 eV) is lower than the hydrogen ionization energy (13.6 eV). 
Thus S$^+$ can form outside the H$^+$ boundary.
We note however that the Balmer jump, sensitive to the electron temperature of H$^+$, yields a temperature that is consistent within $1.5\sigma$ of the S$^+$ temperature. Thus we would not expect a significant change in our measured N$^+$ abundance from this effect. 
Additionally, we calculated the N abundance using the upper limit on $T_e(\textrm{N}^+) < 9000$ K, and find that this is consistent with the abundance measured using $T_e(\textrm{S}^+)$.

\subsubsection{Sulfur Abundance}

Our spectrum contains the temperature sensitive lines for both S$^+$ and S$^{++}$, which we use to calculate the abundance of the lower ionization states of sulfur. 
We use the direct S$^+$ and S$^{++}$ temperatures to measure the abundances of these sulfur ions. 

We do not have a measurement of a higher ionization state of sulfur, so we must correct for the unobserved high ionization sulfur.
There may be a small contribution from S$^{+3}$ ($34.86-47.22$ eV ionization energy).
We use the ionization correction factor (ICF) of \cite{Izotov06}, specifically the ``low-Z" component of their Equation 20, to make this correction. 
We find that the ICF is consistent with 1 within uncertainties. 
As a consistency check, we calculate the sulfur ICF using the corrections reported in \cite{Amayo21}. We find that the ICF is still consistent with 1 within uncertainties.

These ICFs indicate that the majority of the sulfur in this H\textsc{ii} region exists in the low and intermediate ionization states. However the presence of higher ionization \arivdbl ~ suggests we might expect a more significant contribution from the higher ionization S$^{+3}$ state. It is possible that the previously described ICFs are insufficient to describe the high ionization of the Sunburst LCE. To further explore this, we consider the 4-zone photoionization models of \cite{Berg21}. Using their \arivdbl/\ariii ~indicator for $\log(U)$ in the higher ionization region gives $\log(U) \sim -1.67$ for the Sunburst LCE, which would indicate an ICF(S) $\sim 2$. This higher ICF yields $\log(\textrm{S}/\textrm{O}) \sim -1.3$, slightly more than $1\sigma$ above our estimate based on the ICF of \cite{Izotov06} reported in Table \ref{tab:results}. A more specific photoionization model for this source would likely be needed to more accurately constrain the contribution of higher ionization states of sulfur, however for our present analysis we conclude that any potential offset is likely contained within our measurement uncertainties.

\subsubsection{Argon Abundance}

We detect both \ariii~ and \arivdbl, allowing us to calculate the ionic abundances of both Ar$^{++}$ and Ar$^{+3}$. 
We correct for unobserved higher ionization states of argon using the ICF in Equation 23 of \cite{Izotov06}, specifically using the ``low-Z" component based on our measured oxygen abundance.
Most of the gas-phase argon is in the two ionization states which we have measured, resulting in an ICF of $1\pm 0.2$ and a decreased systematic uncertainty on the total argon abundance. 
The ICF of \cite{Amayo21} provides a similar conclusion that we are observing most of the Ar in the nebula, as the ICF is again consistent with 1 within uncertainties.

The ionization energy range of Ar$^{++}$ does not neatly align with any ion for which we have a temperature measurement; it lies between the S$^{++}$ and O$^{++}$ zones. 
The overlap with S$^{++}$ is slightly greater, so we choose to use $T_e(\textrm{S}^{++})$ when calculating Ar$^{++}$/H$^+$.
We did test the effect of using the O$^{++}$ temperature, and found that the resulting Ar$^{++}$/H$^+$ ionic abundance is lower by a factor $\sim 2$.
Ar$^{+3}$ better aligns with O$^{++}$, so we use $T_e(\textrm{O}^{++})$ when calculating Ar$^{3+}$/H$^+$.

\subsubsection{Neon Abundance}

We only detect neon in the Ne$^{++}$ state, requiring a correction for unobserved ionization states. 
We again use the ICF of \cite{Izotov06} for this correction, and check for consistency using the ICF of \cite{Amayo21}. 
The neon abundance and ICFs are reported in  \autoref{tab:results}.
The two ICF estimates are consistent within uncertainties.
We use the electron temperature from O$^{++}$ to calculate the neon ionic abundance.
While we have a direct measurement of $T_e(\textrm{Ne}^{++})$, this temperature has a relatively large uncertainty (owing to the faintness of the \neiiiauroral~line). 
Additionally, it is consistent within $1\sigma$ uncertainties of $T_e(\textrm{O}^{++})$. 
Thus using the doubly ionized oxygen temperature reduces the statistical uncertainty on the neon abundance measurement without introducing any additional systematic uncertainty, as the two temperatures have been found to be consistent.

We find that the ionization correction for neon is $1.1 \pm 0.1$, indicating that the majority of the neon is contained within the measured Ne$^{++}$ state.

\subsubsection{Iron Abundance}
We calculate the gas-phase iron abundance based on the strength of the nebular \feiii ~line.
To account for the unobserved ionization states of iron, we use the ICF from Equation 24 of \cite{Izotov06}.
No Fe ICF is reported in \cite{Amayo21}.
The ionization energy of Fe$^{++}$ (16.2--30.7 eV) overlaps both the low and intermediate ionization zones. We choose to use the low-ionization temperature from S$^+$ for the iron abundance calculation, though the final result would not change significantly if we instead used the S$^{++}$ temperature. 

The iron abundance here relies on only a single ionization state, and as such the ICF ($6 \pm 2$) is large. This abundance is thus subject to additional systematic uncertainties. 
While we detect \feii, this line is strongly affected by flourescence and is thus not suitable for abundance determinations. 
The \feiib ~line appears to be partially detected, however it straddles the detector gap in the NIRSpec IFS data so we cannot measure a reliable flux. 
We do not detect any higher-ionization iron lines, though we may expect some contribution from Fe$^{+3}$, which has similar ionization energy to O$^{++}$.

\cite{Berg21} calculate Fe ICFs as a function of ionization parameter using photoionization models of two nearby extreme emission line galaxies. Their ICFs have a strong dependence on ionization parameter when only Fe$^{++}$ is observed, as is the case for the Sunburst LCE. Using our estimate of $\log(U) \sim -1.67$ based on the \arivdbl/\ariii ~indicator from \cite{Berg21} along with their Fe ICF, we find that the correction could be as high as $\sim 15$. We therefore encourage caution be used in interpreting our reported iron abundance.

\subsection{Helium Abundance}

We measure a variety of He\textsc{i} lines that allow us to calculate the helium abundance, which is needed for the Balmer jump temperature calculation. 
While we detect \heiiopt ~emission, we only see evidence of the broad stellar component, without significant contribution from a narrow nebular component. 
We therefore assume that the He$^{++}/$H$^+$ abundance is small relative to He$^{+}/$H$^+$, and thus He/H = He$^{+}/$H$^+$. 

The He\textsc{i} ionization energy is most similar to that of O$^{++}$, so we use $T_e(\textrm{O}^{++})$ when calculating the helium abundance. 
We have a number of He\textsc{i} lines available to calculate the ionic abundance. 
We find that the various lines give similar abundance results (reported in Table \ref{tab:results}. For our final value, we use an error-weighted average of the abundance measurements from each line. 

\section{Abundance Pattern of an H\textsc{ii} Region at Redshift 2.37}\label{sec:discussion}

\begin{table}
\centering
\caption{Temperature, Density, and Abundance Measurements from the Ly-C Leaking Cluster}
\label{tab:results}
\begin{tabular}{c c}
\hline \hline
Property & Measurement \\
\hline
$T_e(\textrm{O}^{++})$ ($10^3$K) & $15.1 \pm 0.5$ \\
$T_e(\textrm{O}^+)$ ($10^3$K) & $28 \pm 3$ \\
$T_e(\textrm{S}^{++})$ ($10^3$K) & $10.3 \pm 0.6$ \\
$T_e(\textrm{S}^+)$ ($10^3$K) & $10 \pm 1$ \\
$T_e(\textrm{N}^+)$ ($10^3$K) & $< 9.0$ \\
$T_e(\textrm{Ne}^{++})$ ($10^3$K) & $15.4 \pm 1.5$ \\
$T_e$(Balmer) ($10^3$K) & $8.2 \pm 0.2$ \\
\hline
$n_e(\textrm{O}^+)$ (cm$^{-3}$) & $1600^{+1500}_{-200}$ \\
$n_e(\textrm{S}^+)$ (cm$^{-3}$) & $1200^{+500}_{-400}$ \\
$n_e(\textrm{Ar}^{+3})$ (cm$^{-3}$) & $8000^{+10000}_{-5000}$ \\
\hline
O$^{+}$/H$^+$ ($\times 10^5$)& $2.0^{+1.1}_{-0.7}$ \\
O$^{++}$/H$^+$ ($\times 10^5$)& $7.3 \pm 1.2$ \\
$12+\log(\textrm{O}/\textrm{H})$ & $7.97 \pm 0.05$ \\
\hline
$\textrm{S}^+/\textrm{H}^+$ ($\times 10^7$) & $2.3 \pm 0.7$ \\
$\textrm{S}^{++}/\textrm{H}^+$ ($\times 10^7$) & $21 \pm 4$ \\
S ICF (I06) & $1.3 \pm 0.3$ \\
S ICF (A21) & $1.1 \pm 0.2$ \\
$12+\log(\textrm{S}/\textrm{H})$ & $6.48 \pm 0.05$ \\
log(S/O) & $-1.49 \pm 0.14$ \\
\hline
$\textrm{N}^+/\textrm{H}^+$ ($\times 10^7$) & $46 \pm 7$ \\
log(N/O) & $-0.65^{+0.16}_{-0.25}$ \\
\hline
$\textrm{Ar}^{++}/\textrm{H}^+$ ($\times 10^7$) & $4.5 \pm 0.9$ \\
$\textrm{Ar}^{+3}/\textrm{H}^+$ ($\times 10^7$) & $0.7 \pm 0.2$ \\
Ar ICF (I06) & $1.0 \pm 0.2$ \\
Ar ICF (A21) & $0.8 \pm 0.2$ \\
$12+\log(\textrm{Ar}/\textrm{H})$ & $5.72 \pm 0.04$ \\
log(Ar/O) & $-2.24 \pm 0.14$ \\
\hline
$\textrm{Ne}^{++}/\textrm{H}^+$ ($\times 10^5$) & $1.5 \pm 0.3$ \\
Ne ICF (I06) & $1.1 \pm 0.1$ \\
Ne ICF (A21) & $ 1.1 \pm 0.3 $ \\
$12+\log(\textrm{Ne}/\textrm{H})$ & $7.21 \pm 0.04$ \\
log(Ne/O) & $-0.66 \pm 0.11$ \\
\hline
$\textrm{Fe}^{++}/\textrm{H}^+$ ($\times 10^7$) & $3.1 \pm 1.7$ \\
Fe ICF (I06) & $6 \pm 2$ \\
$12+\log(\textrm{Fe}/\textrm{H})$ & $6.3 \pm 0.12$ \\
log(Fe/O) & $-1.7 \pm 0.3$ \\
\hline 
$\textrm{He}^{+}/\textrm{H}^+ (\lambda4471)$ & $0.08 \pm 0.02$ \\
$\textrm{He}^{+}/\textrm{H}^+ (\lambda5876)$ & $0.09 \pm 0.01$ \\
$\textrm{He}^{+}/\textrm{H}^+ (\lambda6678)$ & $0.09 \pm 0.04$ \\
Mean $\textrm{He}^{+}/\textrm{H}^+$ & $0.08 \pm 0.02$ \\
$12+\log(\textrm{He}/\textrm{H})$ & $10.93 \pm 0.09$ \\
\hline

\end{tabular}
\tablecomments{ICFs are reported using the corrections of \cite{Izotov06} (labeled as I06) and using the corrections of \cite{Amayo21} (labeled as A21) where applicable. Abundances are calculated using the \cite{Izotov06} ICFs for consistency across all measured elements.
}
\end{table}

We have detected the auroral lines \siiauroraldbl, \oiiiauroral, \siiiauroral, and \oiiauroraldbl, allowing us to measure the electron temperature in the low, intermediate, and high ionization zones of the individual H\textsc{ii} region of the Sunburst LCE. 
The resulting temperatures are reported in Table \ref{tab:results}, and plotted in Figure \ref{fig:temps} alongside temperature measurements from local H\textsc{ii} regions from the CHAOS project \citep{Berg15,Croxall15,Croxall16,Berg20}.
We do not detect \niiauroral, so we set a $3\sigma$ upper limit on the N$^+$ temperature. 
As discussed in Section \ref{sec:tempdiscuss}, we see a gradient in temperatures between the high-ionization regions at the center of the nebula and the lower-ionization regions towards the edge of the nebula. This gradient is likely driven by the strong radiation produced by the population of massive stars in the central star cluster in this region.

These electron temperatures allow us to calculate abundances of multiple species via the direct method. 
The resulting ionic abundance, ICFs, and abundances relative to oxygen are reported in Table \ref{tab:results}. 
Our spectrum includes emission lines from the dominant ionization states of oxygen, nitrogen, sulfur, argon, and neon, minimizing the uncertainties propagated to our abundances from ionization correction factors. 

This combination of directly measured electron temperatures for each of the low, intermediate, and high ionization zones, plus detection of the relevant emission lines for the primary states of multiple elements, enables the same quality of abundance measurements as has been used in nearby galaxies. 
We can therefore directly compare the gas-phase abundances of the H\textsc{ii} region surrounding this young star cluster at $z=2.37$ to abundances measured in nearby H\textsc{ii} regions. 
This is the first time chemical abundance patterns have been measured to this level of precision in an individual H\textsc{ii} region beyond the local universe. 

\begin{figure*}
    \centering
    \includegraphics[width=0.9\textwidth]{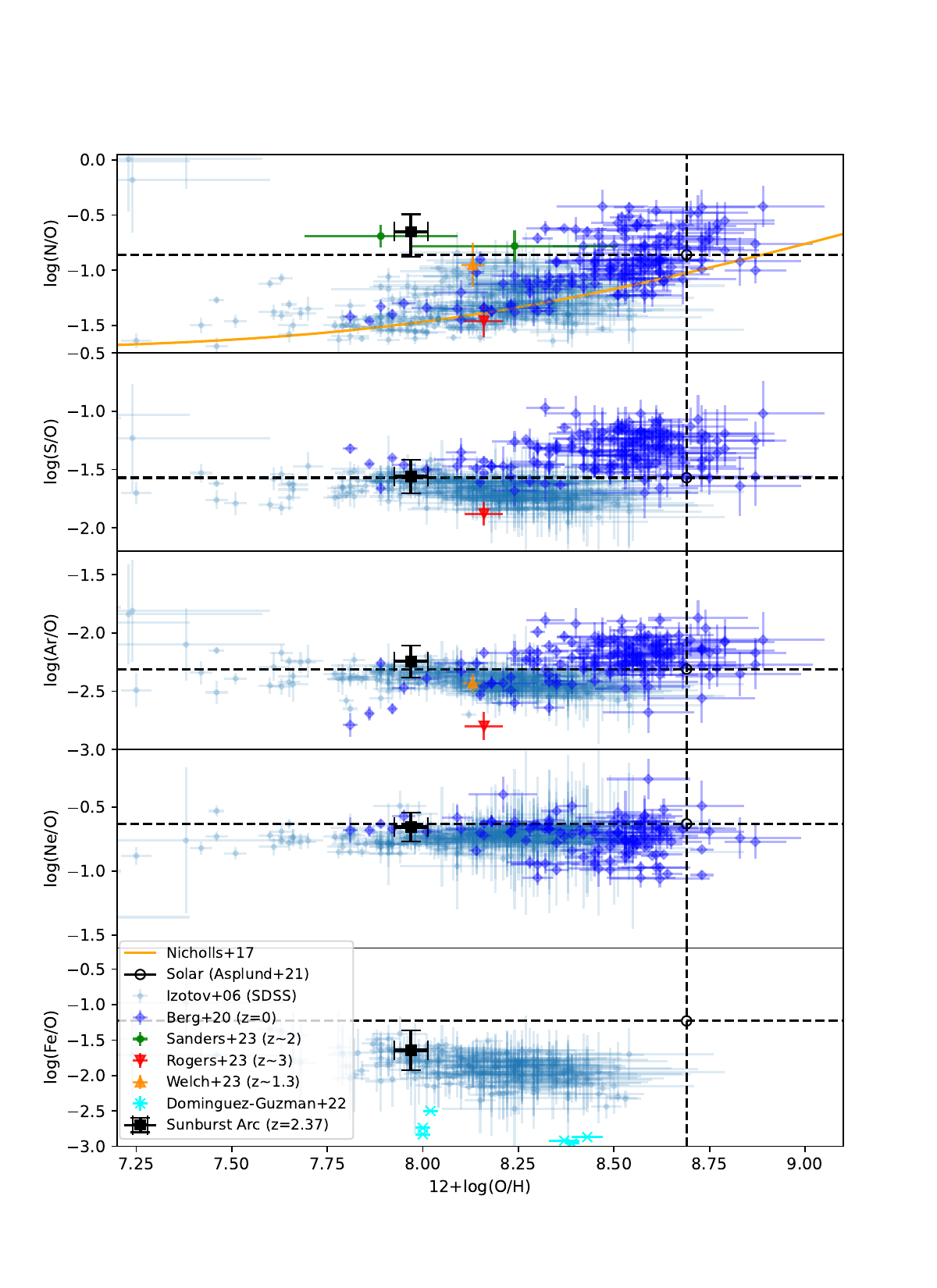}
    \caption{Abundances of the LyC leaking clump in the Sunburst Arc, compared to local H\textsc{ii} regions \citep{Berg20}, low-redshift galaxies \citep{Izotov06}, and other galaxies at Cosmic Noon \citep{Rogers24,Sanders23_keck,Welch24}. Solar abundances from \cite{Asplund21} are shown as black dashed lines. We find an enhanced nitrogen abundance, while other elements are consistent with local H\textsc{ii} regions and low-redshift star-forming galaxies.}
    \label{fig:abunds}
\end{figure*}

We plot the abundances of nitrogen, sulfur, argon, neon, and iron relative to oxygen in Figure \ref{fig:abunds}. 
We compare our abundances to measurements from local H\textsc{ii} regions from CHAOS \citep{Berg15,Croxall15,Croxall16,Berg20}, low-redshift galaxies from SDSS \citep{Izotov06}, and other galaxies measured at cosmic noon \citep{Sanders23_keck,Rogers24,Welch24}.
Solar abundances from \citet{Asplund21} are also marked. 
We find that our measured abundances of sulfur, argon, and neon are all consistent with solar abundance patterns. 
Additionally, these elements are consistent with local H\textsc{ii} regions, indicating the enrichment of alpha elements in the LCE is following standard patterns, in contrast with the low Ar/O abundance for a $z\sim 3$ galaxy reported in \cite{Rogers24}. 
Our nitrogen abundance is high compared to local H\textsc{ii} regions, similar to the $z\sim 2$ galaxies reported in \cite{Sanders23_keck}. 
We discuss this enhanced nitrogen abundance in the following section. 
We find an iron abundance that is slightly below the solar value, however it matches the Fe/O abundances of low-redshift galaxies with similar oxygen abundances \citep{Izotov06}. 
Our measured Fe/O is still higher than what has been observed in the Magellanic Clouds \citep{DominguezGuzman22}, however our measured dust $E(B-V)$ is also lower, indicating that there is less dust available onto which the iron can condense.

\subsection{Nitrogen Enrichment from Wolf-Rayet Stars}

The oxygen abundance of the Sunburst LCE places it just below the point at which measurements of local H\textsc{ii} regions have begun to find significant contribution from secondary nitrogen production. 
We might therefore expect it to only be affected by primary nitrogen production, placing it along the N/O plateau for low-metallicity galaxies and H\textsc{ii} regions (e.g. within uncertainties of the fit to stellar abundance data from \cite{Nicholls17}). 
The elevated N/O measured for the LCE is in tension with this interpretation, indicating another source of nitrogen enrichment is driving the abundance. 

Elevated nitrogen abundances have been observed previously in low-redshift galaxies \citep[e.g.,][]{Brinchmann2008,Amorin2010,Berg11}, galaxies at cosmic noon \citep[e.g.,][]{Masters14,Sanders23_keck}, and more recently in high redshift galaxies \citep{Cameron23,Senchyna23,Marquez-Chavez24,Castellano24}. 
The elevated nitrogen has often been attributed to nitrogen enrichment from Wolf-Rayet stars \citep{Henry2000,Kobayashi24}, though other populations such as AGB stars \citep{Dantona23} and supermassive stars \citep{Charbonnel23} have also been hypothesized as drivers of N enhancement.
It has also been suggested that inflows of low-metallicity gas can dilute the oxygen abundance, driving increases in the N/O ratio without requiring enrichment from Wolf-Rayet stars \citep[e.g.,][]{Koppen2005,Amorin2010,Andrews2013}.

A companion paper utilizing the same stacked spectrum determined that the LCE contains spectral signatures of Wolf-Rayet stars \citep{RivThor24}. 
Previous analyses of galaxies showing WR signatures at low redshift have found that abundances of alpha elements (Ar \& Ne) relative to oxygen were not elevated \citep{Brinchmann2008}. 
Recent theoretical analysis of WR abundance patterns from \cite{Kobayashi24} find a similar result, that only the nitrogen abundance will change significantly with enrichment from WR stars. 
We see a similar abundance pattern here, with only N/O elevated relative to similarly low-O/H galaxies and H\textsc{ii} regions. 
The similar abundance pattern could be indicative of WR stars driving the elevated N/O in the Sunburst LCE. 

Earlier studies have suggested that the WR stellar winds driving the enhanced nitrogen abundance would also carry significant helium \citep{Kobulnicky1997}. 
We find a helium abundance $12+\log(\textrm{He}/\textrm{H}) = 10.93 \pm 0.09$ that is consistent with solar \citep[$12+\log(\textrm{He}/\textrm{H})_{\odot} = 10.914 \pm 0.013$,][]{Asplund21}. 
\cite{Brinchmann2008} also find no evidence for elevated helium abundance in their WR galaxy sample, yet still conclude that the elevated N/O is likely due to WR stellar winds. 
We thus make the same conclusion here, that WR stellar winds are driving increased N/O abundance in the Sunburst LCE.
A recent analysis of several high-redshift ($z\sim 6$) galaxies with elevated N/O also found evidence for elevated He/H \citep{Yanagisawa24}. 
These authors suggest that elevated densities ($n_e \sim 10^3-10^4$) could be related to the high helium abundance. We measure a similar density ($n_e \sim 1200-4500$) but see no increase in helium abundance.

The type of WR stars present can also influence the measured chemical abundances. 
Analyses of wind abundances in WN and WC stars found that WNs exhibit fairly typical neon abundance, while WC stars show significantly elevated levels of neon \citep{Smith05,Ignace07}.
We see no evidence of enhanced neon abundance here, which could be indicative of a majority of the WR stars present being WN-type. 
In exploratory fitting  by \cite{RivThor24}, the best-fitting BPASS model had only 13\% of the WR stars of WC-type, offering some support to this hypothesis. 
However, more detailed analysis would be required to constrain the relative quantities of WC and WN stars. 

Very massive stars (VMS, $M>100M_{\odot}$) have been suggested to be core H-burning WN stars \citep[WNh,][]{Vink23}. The stellar population fits of \cite{RivThor24} suggest $\sim 70$\% of the WR stars in the Sunburst LCE are WNh. Previous analyses have also suggested the Sunburst LCE contains VMS \citep{Mestric23}. Estimates of nucleosynthetic yields from VMS suggest that these stars could contribute significantly to the overabundance of N \citep{Vink23}. Models of VMS also indicate that these stars produce additional Ne compared to regular massive stars \citep{Higgins23}. Our measured Ne/O being consistent with solar could be in mild tension with the VMS yields, however we note that \cite{Higgins23} find strong variation in the Ne production relative to O with changing mass. The IMF could alter the total yields. We therefore cannot rule out contributions from VMS in the Sunburst LCE.

A previous analysis of the LCE using rest-frame UV spectra also measured a high N/O abundance \citep{Pascale23}. 
Their measurement ($\log(\textrm{N}/\textrm{O}) = -0.21 \pm 0.1$) is higher than our direct measurement by $\sim 0.3$ dex, though the two are still consistent within $2\sigma$.
\cite{Pascale23} hypothesize that the nitrogen enrichment is most enhanced in the high-density, high-ionization regions of the cluster. 
Our measurement using N$^+$ being slightly lower than their measurement based on N$^{++}$ would be consistent with this picture. 
However our measurement is significantly higher than the low-density N/O limit ($\log(\textrm{N}/\textrm{O}) < -1.31$) calculated by \cite{Pascale23} based on the non-detection of the \niidbl ~lines.

Our measurements of the abundance pattern of the Sunburst LCE disfavor the supermassive star ($M>1000M_{\odot}$) and AGB star models that have been suggested to explain elevated nitrogen abundances in high redshift galaxies and proto-globular clusters. The AGB model suggested by \cite{Dantona23} requires a longer timescale for nitrogen enrichment ($\sim 100$ Myr), which is incompatible with the young stellar population of the Sunburst LCE star cluster \citep[$\sim 3-4$ Myr,][]{Chisholm19,RivThor24}. The supermassive star model better matches the observed age of the central star cluster; the short lifetimes of these supermassive stars would allow them to meaningfully alter the nebular abundances in just a few million years. However the supermassive star enrichment model of \cite{Charbonnel23} predicts an elevated Ne/O abundance ratio, with the most metal-poor objects showing the largest Ne/O increases due to the strong depletion of O in this model. Our measured Ne/O being consistent with solar appears to be in tension with this model prediction. Additionally, the earlier supermassive star model of \cite{Gieles18} predicts modest enhancement of He in proto-globular cluster environments with a population of supermassive stars. Our measured He/H being consistent with solar is again in mild tension with this prediction, however we note that the models of \cite{Gieles18} do allow for enhanced N with very low levels of He enrichment. Thus while our measurements appear to be in tension with the supermassive star model, we cannot conclusively rule it out based on our abundance measurements.

\section{Conclusions} \label{sec:conclusion}

We have measured direct gas-phase chemical abundances in the stacked spectrum of the Lyman-continuum emitting star cluster in the Sunburst Arc. 
We detect the \siiauroraldbl, \oiiiauroral, \siiiauroral, and \oiiauroraldbl ~lines, allowing measurement of the electron temperature in the low, intermediate, and high ionization zones of the nebular gas. 
With these multiple temperature measurements, we measure direct abundances of oxygen, nitrogen, sulfur, argon, neon, and iron, following best practices established in nearby galaxies from the CHAOS program. 
These measurements enable a direct comparison between well-studied nearby H\textsc{ii} regions and an H\textsc{ii} region at Cosmic Noon ($z=2.37$).

These high-quality abundance measurements enable detailed studies of the origins of abundance patterns in distant galaxies and H\textsc{ii} regions. 
In the Sunburst LCE, the combination of these abundance measurements and the detection of Wolf-Rayet features \citep{RivThor24} allow us to conclude that nitrogen-rich Wolf-Rayet stars are responsible for increasing the nitrogen abundance. 

Further high-quality abundance measurements of galaxies and galactic substructures at Cosmic Noon and beyond will continue to drive our understanding of how galaxies and their star clusters built up the chemical elements. 
For example, further exploration of both temperature and abundance variation in individual H\textsc{ii} regions in a larger sample of gravitationally lensed galaxies could constrain globular cluster formation models, recalibrate empirical temperature relations for high-$z$ studies, and examine abundance gradients to constrain galaxy formation models. This study could thus be seen as a proof of concept for future work characterizing abundances of lensed H\textsc{ii} regions.

\begin{acknowledgments}
The authors wish to dedicate this paper to the memory of Prof. Liese Van Zee, a pioneer of chemical abundance measurements in galaxies. 
We thank the referee for their insightful comments which have strengthened this paper.
This work was based on observations taken with the NASA/ESA/CSA JWST.
JWST is operated by the Space Telescope Science Institute under the management of the Association of Universities for Research in Astronomy, Inc., under NASA contract no. NAS 5-03127.
These observations were taken as part of JWST GO program 2555, support for which was provided by NASA through a grant from the Space Telescope Science Institute, which is operated by the Association of Universities for Research in Astronomy, Inc., under NASA contract NAS 5-03127.
The JWST data analyzed in this paper were obtained from the Mikulski Archive for Space Telescopes (MAST) at the Space Telescope Science Institute. The specific observations analyzed can be accessed via \dataset[doi:10.17909/c6fa-7714
]{http://dx.doi.org/10.17909/c6fa-7714}.
BW is supported by NASA under award number 80GSFC21M0002. ER-T is supported by the Swedish Research Council grant
No. 2022-04805\_VR.
\end{acknowledgments}

%

\vspace{5mm}
\facilities{JWST}


\software{Astropy \citep{Astropy13,Astropy18,Astropy22}, 
          Matplotlib \citep{Matplotlib_Hunter07},
          Scipy \citep{2020SciPy-NMeth},
          emcee \citep{ForemanMackey13_emcee},
          PyNeb \citep{pyneb_luridiana15}
          }



\bibliography{bib}{}
\bibliographystyle{aasjournal}



\end{document}